# Revealing the First Stellar and Supermassive Black Holes to *EXIST*

J. Grindlay[a], J. Bloom[b], P. Coppi[c], A. Soderberg[a], J. Hong[a], H. Moseley[d], S. Barthelmy[d], G. Tagliaferri[e], G. Ghisellini[e], R. Della Ceca, P. Ubertini[f]

[a]Harvard-Smithsonian CfA, 60 Garden St., Cambridge, MA, [b]UCB, Berkeley, CA, [c]Yale Univ., New Haven, CT, [d]NASA-GSFC, Greenbelt, MD, [e]INAF-OAB, Merate, Italy, [f]INAF-IASF, Rome, Italy

**Abstract.** The epochs of origin of the first stars and galaxies, and subsequent growth of the first supermassive black holes, are among the most fundamental questions. Observations of the highest redshift Gamma-Ray Bursts (GRBs) will be the most compelling *in situ* probe of the history of initial star formation and consequent epoch of reionization if their prompt and precise detection can be followed immediately by sensitive near-IR imaging and spectroscopy. Blazars are the persistent analogs of GRBs and for the same reason (beaming) can be observed at highest redshifts where they might best trace the high accretion rate-driven jets and growth of supermassive black holes in galaxies. The proposed *EXIST* mission can uniquely probe these questions, and many others, given its unparalled combination of sensitivity and spatial-spectral-temporal coverage and resolution. Here we provide a brief summary of the mission design, key science objectives, mission plan and readiness for *EXIST*, as proposed to Astro2010.

## INTRODUCTION

Wide-field imaging full-sky surveys, with rapid cadence, are needed for the discovery and study of rare classes of objects, which can provide unique constraints on the Early Universe. Foremost among these are Gamma-Ray Bursts (GRBs) which, with the discovery of GRB090423 at $z = 8.3$ (Salvaterra et al 2009, Tanvir et al 2009), have rapidly surpassed luminous quasars as the most distant objects with spectroscopic redshifts. As discussed by Bromm and Loeb (2007, and references therein), the non-thermal afterglow emission (fading with power law decay typically as $t^{-1.3}$) is detectable in its decay earlier (and thus brighter) by the redshift time dilation factor $1 + z$. Thus GRBs at high z are the optimum probes of the Early Universe provided that spectra of their luminous afterglows can be obtained promptly at wavelengths longward of their spectral cutoffs at redshifted Ly-α, or at $\geq 1.216\mu m[(1 + z)/10]$. Just as the relativistic beamed hard X-ray luminosity of GRBs enables their initial prompt detection by wide-field coded aperture imaging telescopes, so does the lower Γ beaming from the variable emission of blazars. Scaling from the detection of blazars out to $z = 3.68$ by the Swift/BAT survey (Ajello et al 2009), which showed that luminous flat spectrum radio quasars (FSRQs) are detected in hard X-rays at larger redshifts than at GeV energies, Ghisellini et al (2010) have shown that these objects may reveal the growth of supermassive black holes (SMBHs) in the early Universe.

In this paper, we first summarize and update our previous description (Grindlay et al 2009) of the proposed *EXIST* mission. We then describe its first two key science objectives -- to explore the early Universe ($z >7$) with **1)** *GRBs* and **2)** *blazars* -- with its unique wide-field hard X-ray (5-600 keV) survey and immediate followup with sensitive broad-band (IR-visible through soft γ-ray) high resolution imaging and spectra. The initial 2y all-sky scanning survey and then 3y pointed study phase will both produce large samples of GRBs and blazars (and AGN of all types) to constrain their physics and evolution. These wide-field surveys also enable our third key science objective: **3)** to discover and study *Transients*, from stellar flares and BHs within the Galaxy, to ULX outbursts in the Local Group, to SGR flares, SNe breakout shocks and tidal disruption flares by quiescent SMBHs out to ~200Mpc. Finally, we mention the mission development and our recent successful first balloon flight test of *ProtoEXIST1*, which demonstrate that *EXIST* could be launched in this decade.

## SUMMARY OF *EXIST*

The Energetic X-ray Imaging Survey Telescope (*EXIST*) was originally selected for a New Concept Study (together with GLAST/*Fermi* and what became Con-X/IXO) as a scanning wide-field hard X-ray (~10-600keV) coded aperture telescope for studying obscured AGN, and secondarily GRBs. It was then proposed to and recommended by the 2000 Decadal Survey as an attached payload of 8m$^2$ of imaging Cd-Zn-T (CZT) detectors in 2 x 4 fixed telescopes on the ISS (Grindlay et al 2001). With the advent in 2003 of NASA's Beyond Einstein Program, *EXIST* became a leading candidate to be the Black Hole Finder Probe, one of three Einstein Probe missions, and was studied for the Beyond Einstein Program Assessment Committee (BEPAC) as a two-instrument (both coded aperture) mission: a High Energy Telescope (HET; 10-600 keV) with 6 m$^2$ of imaging CZT in 19 separately shielded sub-telescopes, and a Low Energy Telescope (LET; 3-20 keV) with 1.3 m$^2$ of imaging Si detectors in 32 separate sub-telescopes (see Fig. 2 of Grindlay et al 2006a). The required mission was massive (9000kg), large (requiring a 5m fairing launcher) and thus expensive; and its GRB and obscured AGN science was limited by having to rely on optical or IR identifications from the ground. After receiving support for an Astrophysics Strategic Mission Concept (ASMC) Study, in preparation for the 2010 Decadal Survey (Astro2010), *EXIST* was radically re-designed to be much smaller, less massive and expensive, and fully autonomous for on-board source identifications.

Fig. 1 shows the overall mission payload, with a single large area (4.5 m$^2$) HET (5 - 600keV) and hybrid mask with 10X finer resolution than *Swift*/BAT, an optical-nIR (0.3 – 2.3μm) 1.1m aperture telescope (IRT) for 4-band imaging and spectroscopy, and a Soft X-ray Imager (SXI) telescope with 0.1 – 10 keV imaging and fast timing. All 3 instruments have significantly finer resolution, broader bands and ≥10$X$ the sensitivity of their counterparts (BAT, UVOT and XRT) on *Swift*. The instrument parameters for all three instruments are given in Grindlay et al (2009), and details are given for the HET instrument by Hong et al (2009) and imaging by Skinner et al (2009); for the IRT by Kutyrev et al (2009) and for the SXI by Tagliaferri et al (2009).

While perhaps outwardly appearing to be a "super-*Swift*", the optimized *EXIST* design (Fig. 1) is conceptually different:
*Scanning vs. pointing:* The orbital scans (Fig. 1) enable both higher sensitivity (by averaging out systematics) and sky coverage (full sky every 3h) for the 2y scanning survey. Followup pointings for GRBs and Transients during the 3y pointed-mode mission phase enables ~full-sky HET coverage each day. Fast readout in the SXI enables photon counting and half-sky surveys in scanning mode, but the readout mode of the H2RG detectors and fine guiding for IRT restrict it to pointing modes (which is ~70% of the 5y mission).
*Deep optical-NIR imaging and spectra:* The 1.1m aperture for the IRT combined with its passively cooled (-30C) optics gives backgrounds at 2 μm that are ~1000X lower than on ground. Thus the IRT is ~10X faster than Keck, imaging to AB =24 in all four bands in just 100s.
*Real-time identifications and redshifts:* With initial HET detection (>5σ)

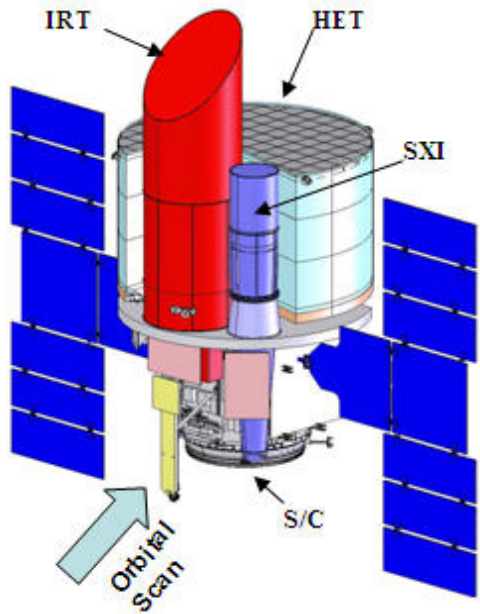

**FIGURE 1.** Wide-field HET (90° x 70°), and narrow-field IRT (5' x 5') and SXI (20') co-aligned on the *EXIST* spacecraft. Zenith pointing 23° North and South of the $i = 15°$ orbital plane on alternate orbits surveys the full sky every ~3h for HET and half-sky every 6mo for SXI during the 2y scanning survey. IRT operates in pointings (~70% of full mission).

positions of <20" (90% conf.), and <150s re-pointings for initial SXI positions of <2", the IRT imaging (0.15" pixels) in 4 bands simultaneously (via 3 dichroics) covering 0.3-0.52-0.9-1.38-2.2μm means that GRB IDs will usually be obvious from comparison with 4' deep images generated on board from stored catalogs, and the object can be moved onto the low-res (R = 30) objective prism slit for a redshift measurement (AB = 23 in 300s) across all 4 bands or onto the high-res (R = 3000) long slit (4") for spectra (AB = 19 in 2ksec) to resolve the damped Ly-α absorption (DLA) out to $z \sim 20$ (for sufficiently bright GRBs) and thus the ionization of the host galaxy vs. IGM and epoch of reionization (McQuinn et al 2009).
*All-timescales, all-time-all-sky imaging:* Because it needs ~100X less telemetry and is also required for scanning, *EXIST* will bring down photon data for HET (and SXI) rather than binned detector images. HET and SXI event data will include all event times (to ~10μsec) as well as energies, position and status so that imaging and spectra can be done on any timescale and energy binning, for any source or sources in any region of sky covered in scanning or pointing. Thus, for example, pulse profiles of X-ray pulsars or lightcurves of AGN could be retrieved over the full mission timescale (subject, of course, to S/N limits per phase bin for pulsars or time bin for lightcurves).
*Serendipitous surveys from ~3y pointing mode:* Finally, during the 3y pointed mission phase for followup of ~30,000 sources discovered in the 2y scanning survey, and continued slew triggers on GRBs, the SXI will cover ~30% of the sky to sensitivity ~1 x $10^{-14}$ cgs and so conduct a *wide*-field, survey for galaxy clusters, obscured AGN and other objects. With exposures ≥2000sec/field, and simultaneous uv-NIR high resolution (0.15") imaging (to AB = 27!) and objective prism spectroscopy partially covered by the IRT for the central 5' x 5' of each SXI field, *EXIST* discoveries would enable/trigger followup studies with very large telescopes (GMT, TMT) and JWST.

## KEY SCIENCE: HIGH-*Z* GRBS AND BLAZARS; TRANSIENTS

Given these powerful (in fact, unique) capabilities, the promise of *EXIST* science is immense and will be (like all "break through" missions) surpassed by the unexpected. We touch on a (very) few highlights for each of the three primary science objectives.

### GRBs: the Long and the Short of it

The primary raison d'etre for the HET+SXI+IRT is to obtain nIR (~1-2.2µm) spectra of Long GRBs produced by the core collapse of massive (≥20-100 $M_\odot$) stars at redshifts $z$ >7. Their continuum afterglows enable absorption line surveys of the host galaxy vs. local IGM to measure or constrain the epoch of reionization (EoR), SFR and metallicity of the Universe vs. $z$. These fundamental questions can best be answered with the in-situ measures of spectra that GRBs can best provide (McQuinn et al 2009) and which require a mission with deep, prompt *nIR spectroscopy from space*. Atmospheric transparency, OH emission, thermal backgrounds, and declining (with time) afterglow emission combine to prevent all but the brightest high-*z* GRBs to be done from the ground (even with Keck or VLT). Simulated IRT spectra for GRB080913 (at $z$ = 6.7), but redshifted to $z$ = 8 and 10, are given in Grindlay et al (2009). Fig. 2 shows GRB lightcurves observed with *Swift*, extrapolated to nIR H, and redshifted to their time-dilated, flux-reduced values if they were at $z$ =12. The 70% with AB ≤19 could have high resolution (R = 3000) spectra measured at the expected IRT sensitivity limit for a 2000s integration (Kutyrev et al 2009). Each would constrain the local EoR and metallicity. GRBs even ~4mag fainter would still have their redshifts measured with the low resolution (R =30) slit or objective prism in 300sec after acquisition in the IRT (or within ≤450s after trigger).

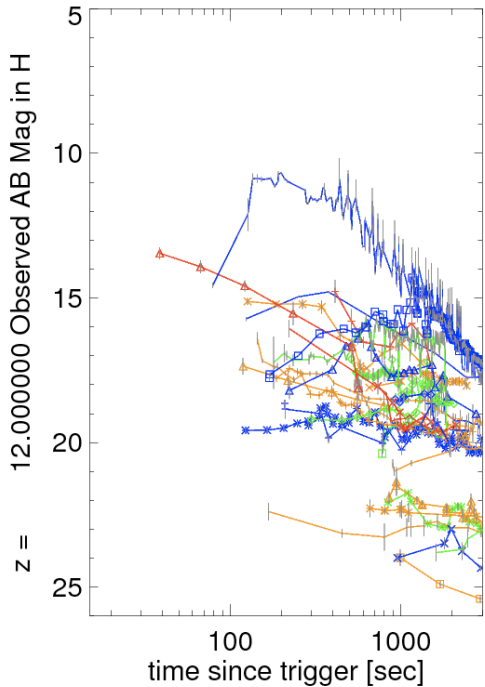

**FIGURE 2.** H-band *Swift* GRB afterglows extrapolated to H band and redshifted to $z$ = 12. IRT redshifts for all could be measured in 300s.

*EXIST* is ≥10X more sensitive to GRBs (particularly high *z*) than *Swift* due to its scanning (larger sky coverage), lower energy threshold (5 vs. 15 keV) and lower backgrounds (active shielding and hybrid mask). The expected GRB rate is ~400-600/y with a redshift distribution (vs. *Swift*) as shown in Fig. 3 as derived by Salvatera (2009) for the comparative mission parameters and assumed SFR vs. *z*. Despite (very) incomplete deep ground based nIR spectra, a second *Swift* GRB at *z* >8 (GRB100205A) is possible (Perley et al 2010). Given that 280 *Swift* GRBs now have optical/IR IDs, this 2/280 = 0.7% fraction is consistent with Fig. 3 and suggests *EXIST* would measure redshifts and spectra for ≥20 – 30 GRBs/y at *z* > 8.

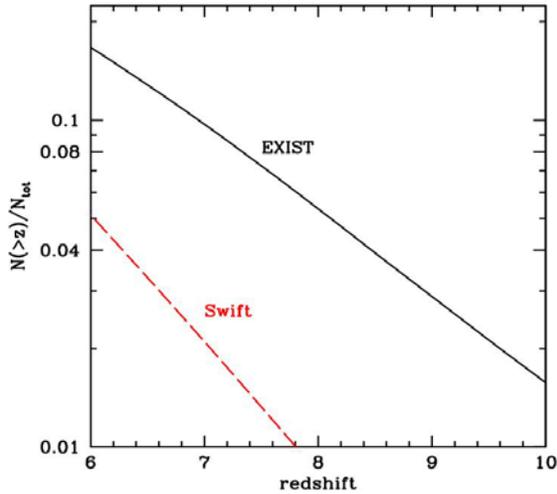

**FIGURE 3.** Expected GRB rates vs. redshift.

The high sensitivity and broader bandwidth will allow definitive studies of GRB physics with high temporal resolution spectra and polarization. *EXIST* will constrain the nature (NS-NS vs. NS-BH mergers?) and hosts of the short hard bursts (SHBs). The NS-NS mergers will be detectable as gravitational wave in-spirals with Advanced LIGO, for which the large and precisely located SHB sample from *EXIST* will enable fundamental measurements (Bloom et al 2009). For those with IRT detections offset from galaxies, the 0.1" positions will enable JWST to search for globular cluster hosts (Grindlay et al 2006b), with AB ~33 and offsets ~2" for a "typical" SHB at *z* = 0.5.

## Blazars and Obscured AGN

As discussed by Ghisellini et al (2010), the blazar (FSRQ) PKS2149-306 detected by BAT at *z* = 2.34 could be detected by *EXIST* out to *z* ~ 8, as shown in Fig. 4. Such objects would be discovered in the HET + SXI scanning survey and have their SED and spectra as well as redshifts measured in the pointed mode followup. The sensitivities of all three instruments are shown, and also that for *Fermi*, which just detects (red "bowtie") the object but would not at larger *z*. The luminosity function derived by Ajello et al from BAT predicts *EXIST* would detect ~15 FSRQ blazars at $z > 7$ with Lx $\geq 2 \times 10^{47}$ erg/s if they exist. Such luminous blazars imply SMBH masses $\geq 10^9$ M$_\odot$ which are challenging to assemble within a Universe age of $\leq 0.8$Gy. Thus the $z$-distribution of the >500-1500 FSRQs that will be identified with *EXIST* at $z > 4$ will provide powerful constraints on the initial growth of SMBHs.

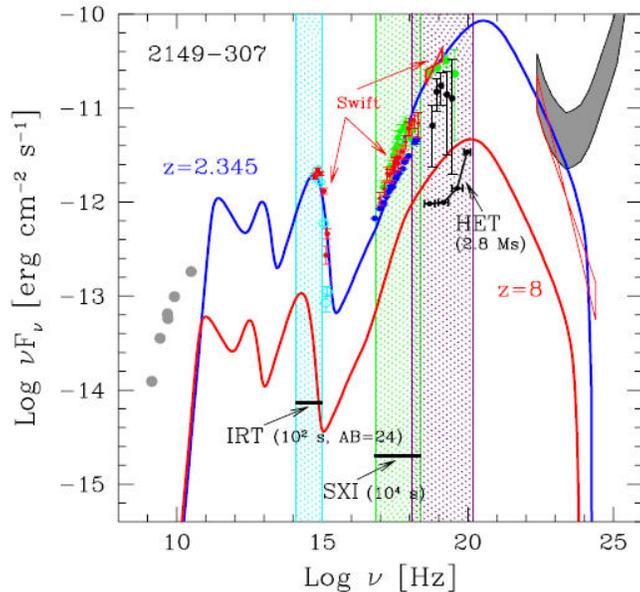

**FIGURE 4.** SED of BAT blazar PKS2149 at *z* = 2.3 is readily detectable at *z* = 8 by all 3 instruments on *EXIST*.

Likewise, for obscured (and Compton thick) AGN, the *EXIST* survey and followup nIR+SXI+HET spectra, redshifts and variability studies will provide an unbiased inventory of their relative numbers, evolution and SMBH mass constraints from variability studies (see Della Ceca et al 2009 and references therein).

## Transients and Time Domain Astronomy (TDA) Surveys

The full sky coverage from both continuous scanning and pointed phase will produce a torrent of transients, from stellar super flares and stellar BH vs. NS X-ray "novae", to SNe breakout shocks (ICECUBE triggers), to tidal disruption of field stars by quiescent SMBHs in galaxies out to ~200Mpc (LISA triggers), as summarized by Soderberg et al (2009). The long-suspected, but never seen population, of *isolated* stellar mass BHs accreting from GMCs in the Galaxy can be discovered by their hard X-ray and nIR signatures. *EXIST* will extend LSST, SKA and TDA to their extremes.

## READY TO *EXIST*

With the successful balloon flight test (October, 2009) of *ProtoEXIST1* employing key technologies for HET (Hong et al 2009) and the forthcoming development of finer pixel CZT imagers derived from *NuSTAR*, and with the flight qualified key components for both IRT (Kutyrev et al 2009) and SXI (Tagliaferri et al 2009), *EXIST* could be launched in 2017. We are grateful for support from NASA grant NNX08AK84G and ASI (Italy) grant I/088/06/0.